\documentclass[9pt,cites,graphicx]{article}
\usepackage{epsfig}
\usepackage{amsmath}
\title{Continuous guided beams of slow and internally cold polar molecules}
\author{Christian Sommer, Laurens D. van Buuren, Michael Motsch,\\ Sebastian Pohle, Josef Bayerl, Pepijn W.H. Pinkse\\ and Gerhard Rempe
\\[3mm] Max-Planck-Institut f{\"u}r Quantenoptik, Hans-Kopfermann-Str. 1,\\ 85748 Garching, Germany. E-mail: pepijn.pinkse@mpq.mpg.de}
\begin{document}
\maketitle
\renewcommand{\thefootnote}{\fnsymbol{footnote}}
\noindent We describe the combination of buffer-gas cooling with electrostatic velocity filtering to produce a high-flux continuous guided beam of internally cold and slow polar molecules.
In a previous paper (L.D. van Buuren {\it et al.}, {\it arXiv:} 0806.2523v1) we presented results on density and state purity for guided beams of ammonia and formaldehyde using an optimized set-up. Here we describe in more detail the technical aspects of the cryogenic source, its operation, and the optimization experiments that we performed to obtain best performance. The versatility of the source is demonstrated by the production of guided beams of different molecular species.

\section{Introduction}
\label{intro}
The successful implementation of laser cooling of atoms in combination with evaporative cooling culminated in the spectacular realization of Bose-Einstein condensation more than a decade ago \cite{Anderson1995,Bradley1995,Davis1995}. A natural next step seems to be the application of similar cooling techniques to molecules. Due to their rich internal structure, molecules can be employed to study 
phenomena which cannot be studied with atoms or are enhanced in molecular experiments \cite{Doyle2004}. Examples are long-range anisotropic electric dipole-dipole interactions in the case of polar molecules, cold collisions, where chemical reactivities are governed by quantum tunneling and resonances \cite{Herschbach1999,Balakrishnan2001} or the measurement of the electron electric dipole moment (EDM) \cite{Hinds1997,Tarbutt2004} in high-precision spectroscopy on heavy dipolar molecules. However, the complexity of the molecules makes it difficult to apply laser cooling techniques, due to the large amount of decay channels inherent to any electronic excitation.

For this reason different approaches to produce cold molecular gases are considered. These include so called indirect methods like photoassociation \cite{Jones2006,Schloeder2001,Kerman2004} or magnetic Feshbach resonances \cite{Kohler2006,Inouye2004,Stan2004} to associate molecules from ultracold atomic samples. As these techniques have the disadvantage of being restricted to a few species and to smaller (mainly dimer) molecules, new methods to create cold samples from naturally given molecules  are explored. Several direct methods have been developed so far: buffer-gas cooling \cite{Weinstein1998}, electric \cite{Bethlem1999,Bethlem2002,vandeMeerakker2008}, magnetic \cite{Narevicius2008,Vanhaecke2007} and optical \cite{Fulton2004,Fulton2006} deceleration, rotating nozzles \cite{Gupta1999}, collisions of molecular beams \cite{Elioff2003} and collisions with moving surfaces \cite{Narevicius2007}, molecules embedded in helium droplets \cite{Stienkemeier2001} as well as velocity filtering by rotating mechanical filters \cite{Deachapunya2008}, and by electrostatic \cite{Rangwala2003,Junglen2004a} or magnetic \cite{Patterson2007} guides.

\begin{figure}[ht]
\begin{center}
\includegraphics[scale=0.9]{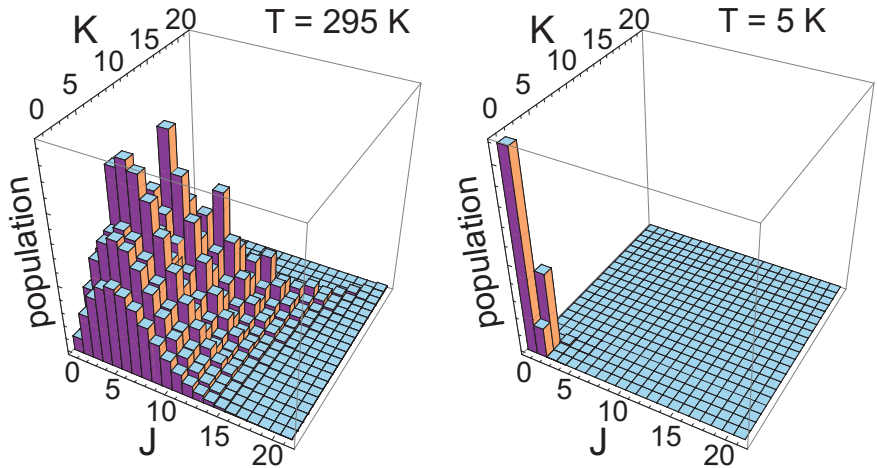}
\caption{Sketches of the state distributions of fully thermalized ND$_3$ ensembles to $295\,$K (left) and to $5\,$K (right). $J$ represents the rotational quantum number with $K$ its projection on the molecular axis.}
\label{fig1}
\end{center}
\end{figure}

Most experiments mentioned above will benefit from large and dense samples of cold molecules. The production of such samples is the main goal of these new direct methods. In our source, we have combined electrostatic velocity filtering and buffer-gas cooling to obtain a continuous beam of dense and internally cold molecules. Warm molecules are introduced into a cryogenic helium buffer gas where all degrees of freedom are cooled by collisions with the helium atoms. As a result the number of populated states are strongly reduced. This is shown in Fig.~\ref{fig1}, where state distributions of deuterated ammonia are presented for fully thermalized ensembles at room temperature and at a temperature of $5\,$K. Slow molecules are extracted out of the buffer-gas environment with an electrostatic quadrupole guide.
In a recent experiment buffer-gas cooling was combined with a magnetic guide consisting of permanent magnets \cite{Patterson2007}. Buffer-gas cooling is a very general cooling method that can be applied to atoms and molecules alike to reach temperatures in the sub-Kelvin regime. In our set-up it is used as a cooling technique that delivers large quantities of slow and internally cold molecules into an electrostatic guide. In combination with laser depletion spectroscopy, applied in the near ultraviolet regime as decribed in \cite{Motsch2007}, the internal cooling was confirmed in our set-up for formaldehyde molecules \cite{vanBuuren2008}. In this paper, we present experiments performed with ND$_{3}$ molecules to optimize and characterize the performance of the cryogenic source. Several parameters of the set-up are varied to maximize the guided flux and internal cooling. To show the generality of the system, measurements with other species like trifluoromethane (CF$_{3}$H) and fluoromethane (CH$_{3}$F) are carried out.

$~$\\

The paper is organized in the following way. The set-up of the cryogenic source for cold polar molecules is presented in Section $2$. In Section $3$ the experimental control and data acquisition system is outlined. Section $4$ presents the method to determine the extracted flux and density by the electrostatic guide. The steps taken to improve the performance of the system are discussed in Section $5$. The guiding of different species is presented in Section $6$ followed by the conclusions and outlook in Section $7$. Additional information about the calibration of the gas densities in the cell is presented in the Appendix.

\section{Experimental Set-up}
\label{exper}
In our experiment molecules from a room temperature source are injected into a cryogenic helium buffer gas via a heated capillary. By collisions with the cold helium gas the internal (rotation) and external (motion) degrees of freedom of the molecules are cooled. \footnote[1]{For ND$_{3}$ the lowest vibrational mode is at $749\,$cm$^{-1}$ \cite{Herzberg}. Therefore, only the ground-vibrational state is populated in these experiments.}
A fraction of the cold molecules is extracted by a quadrupole guide \cite{Rangwala2003,Junglen2004a} to an ultrahigh vacuum environment at room temperature, where they are available for further experiments ({\it e.g.} \cite{Rieger2005,Willitsch2008}).

The technical implementation of the cryogenic source is presented in detail in this section, accompanied by several drawings of the set-up. For more background information about buffer-gas cooling, references \cite{Weinstein1998,Maxwell2005} can be consulted, whereas more information on velocity filtering can be found in \cite{Rangwala2003,Junglen2004a,Motsch2008}.

As shown in Fig.~\ref{fig2}, the cryogenic source is located in a stainless steel vacuum vessel (height $\sim 60\,$cm and diameter $\sim 40\,$cm). It mainly consists of a cryogenic cooler to which two radiation shields and a gold-coated copper buffer-gas cell are mounted. The cell can be filled with helium gas via a gas supply line, which is thermally connected to the $5\,$K cooling stage of the refrigerator. Molecular gas can be introduced into the buffer-gas cell from a thermally isolated gas supply line. To extract the molecules from the buffer-gas cell, an aperture is made in the cell just in front of the electric quadrupole guide. Except for the turbo-molecular pump ($\sim 550\,$l/s), all components are mounted to the top flange to allow for easy servicing of the system.

\subsection{Cryogenic System}
\label{cryo}
\begin{figure}[t]
\begin{center}
\includegraphics[scale=0.8]{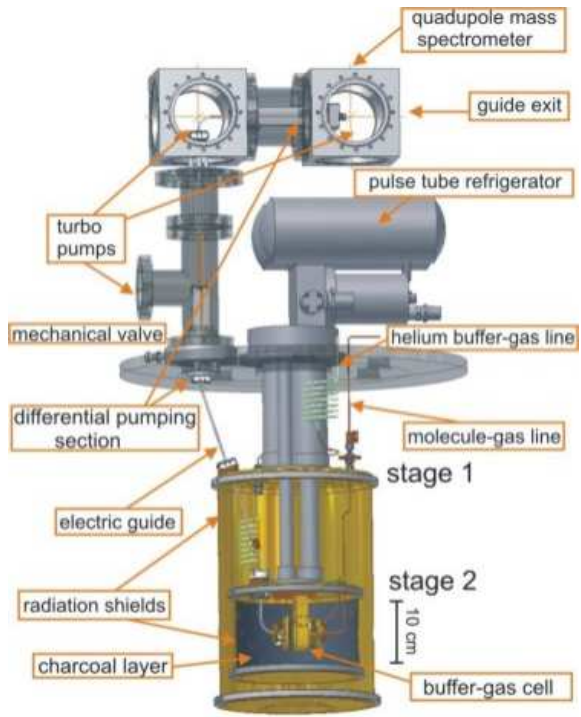}
\caption{Experimental set-up. The inner assembly is mounted directly to the refrigerator. With two capillaries, helium and molecular gases are injected into the buffer-gas cell. While the helium line is thermally connected to the first and the second stage of the pulse tube cooler, the molecular line is thermally isolated from these stages. With an electric quadrupole guide molecules are extracted from the buffer-gas at the exit aperture of the buffer-gas cell and transported to a high-vacuum environment, positioned above the vessel. This part is separated from the vessel by differential pumping stages. During warm up, a mechanical valve can isolate these two regions completely.}
\label{fig2}
\end{center}
\end{figure}
To cool the buffer-gas cell to cryogenic temperature, we use a commercially available pulse-tube cooler (CRYOMECH PT410). For our purposes, a pulse-tube cooler is advantageous compared to cooling schemes based on cryogenic liquids. It is a closed system that delivers enough power to reach cryogenic temperatures in our set-up (T$_{min} \sim 3\,$K) while vibrations are small (in the range of $\sim 20\, \mu$m). Other advantages are the fast cool down and warm up times ($\sim 2\,$hours cool down; $\sim 8\,$hours warm up). The two-stage design of the pulse-tube cooler allows for a configuration, where the cryogenic cell can be insulated from thermal radiation of the vacuum vessel as shown in Fig.~\ref{fig2}. The first (upper) stage can maintain a temperature of $\sim 45\,$K with a heat load of $40\,$W and the second (lower) stage can reach temperatures less than $\sim 4\,$K if the heat load is reduced to less than $1\,$W. Therefore, round gold-plated copper radiation shields are installed on each stage of the cryogenic cooler to reduce the heat loads on both stages. To monitor the temperatures in our set-up we use silicon diode temperature sensors (LAKESHORE DT$470$) at various positions on the two stages. The sensors allow temperature measurements in the range of $1.4\,$K to $475\,$K with an accuracy of $\pm 0.5\,$K at low temperatures.

\begin{figure}[t]
\begin{center}
\includegraphics[scale=0.9]{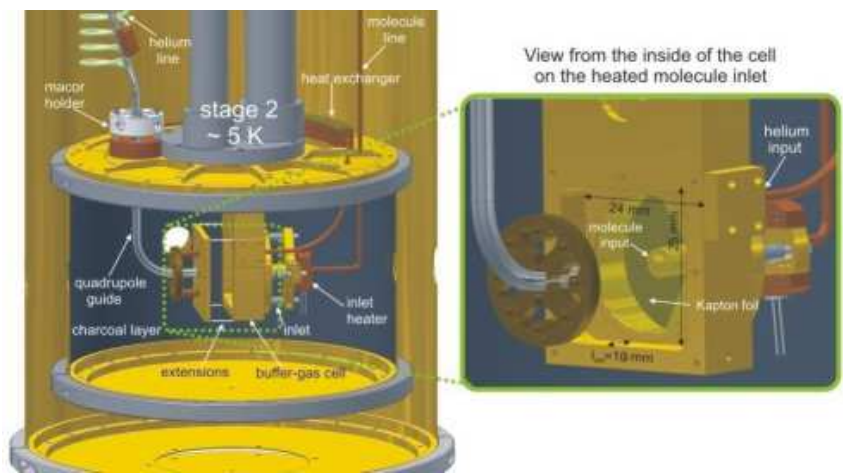}
\caption{The buffer-gas cell in the cryogenic environment. The helium line enters the copper cell from the side.  The heated molecular line is connected to a special inlet assembly, which has a bad thermal connection to the buffer-gas cell. This allows for the heating of the inlet up to room temperature without affecting the temperature of the buffer-gas cell much. The length of the cell can be varied by placing $1\,$cm long copper extensions in between the main frame of the cell and the exit plate. Outside of the buffer-gas cell, a charcoal coated surface pumps away the helium and non-guided molecules. Note that the exit plate of the cell is omitted in the right drawing.}
\label{fig3}
\end{center}
\end{figure}

The inside of the radiation shield mounted to the second stage is covered with a layer of activated charcoal (Fig.~\ref{fig2} and Fig.~\ref{fig3}). At cryogenic temperatures activated charcoal acts as a pump for helium. Large pumping speeds can be obtained due to its large effective area. Because of its simplicity it represents the optimal solution for our experiment. It is applied in our set-up to keep the pressure within the radiation shields as low as possible. From test experiments we concluded that for pressures lower than $\sim 10^{-6}\,$mbar collisions of guided molecules with background gas are negligible. A reasonable vacuum is also required to prevent discharges between the guide electrodes. We used coconut-based granular activated carbon (Chemviron Carbon $207$C). A surface of roughly $1000\,$cm$^{2}$ is covered with a layer of this material, resulting in a pumping speed of $\sim 5000\,$l/s at cryogenic temperatures. The gas (mainly helium) adsorbed by the charcoal is released during warm up of the cryogenic system after which it is pumped out of the vacuum vessel by the turbo-molecular pump.

\subsection{Buffer-Gas Cell and Gas Supply System}
\label{cell}
To bring the molecular gas as well as the helium gas into the cryogenic environment, a  feed through for two $6\,$mm wide gas lines is installed on the top plate of the vessel (see Fig.~\ref{fig2}). The buffer-gas line is attached to one of the feed through lines. The total buffer-gas line consists of four connected segments. For the first segment of the buffer-gas line, a stainless steel pipe with a $3\,$mm outer and $1\,$mm inner diameter is used and connected to the feed through into the vacuum chamber. The stainless steel line contains several windings to increase the length, thereby reducing the heat conductance of the gas line. Since the buffer gas needs to be cooled down on its way to the cell, the stainless steel segment of the line is connected to a subsequent copper line segment with equal dimensions. This copper part of the line is tightly connected to the first stage of the cryocooler for pre-cooling of the helium gas to 35\,K. It then enters into the volume enclosed by the outer radiation shield, where it is connected to a stainless steel pipe with equal dimensions for thermal insulation. To cool the helium gas to 5\,K, the stainless steel gas line is passed over into a copper segment, which is mounted to the second stage of the cryocooler with good thermal contact. From there, the helium gas line enters the inner radiation shield and is finally connected to the buffer-gas cell from the side (see Fig.~\ref{fig3}). It is sealed with a thin indium foil and clamped via a small copper block to the cell wall.

The molecular gas line is heated to maintain a sufficiently high vapor pressure and to avoid freezing. For the molecules employed in this work, we used temperatures above $140\,$K (for ND$_{3}$ above $180\,$K). In most of the presented data in this paper, the molecular input capillary is kept at $295\,$K. Electric heaters and sensors (PT $100$ and DT$470$) are placed at various positions to keep the line temperature fixed and prevent cold spots. To limit heat loads on the cell, the molecular gas line must be thermally disconnected from the cooling stages and the buffer-gas cell. The line is made of a $3\,$mm wide copper pipe that goes from the feed through to the second stage, where it is inserted into a gold-plated copper inlet with an inner diameter of $1\,$mm. A small Viton O-ring is used to seal the transition between the gas line and the inlet. To avoid heating by the molecular gas line, the inlet frame is clamped to the cell via small glass balls to minimize the contact surface. The conical inlet tip is glued on a $25\,\mu$m thick polyimide (Kapton) foil with a $1\,$mm opening. The Kapton foil forms the inlet front wall of the buffer-gas cell and reduces the heat load from the inlet to the cooper cell due to its low thermal conductance. This guarantees that the heat conductance from the inlet to the buffer-gas cell is low enough to maintain cryogenic temperatures, even for a gas line temperature of $\sim 300\,$K.

The flow of the molecular and helium gas can be controlled via a gas handling system. To sensitively regulate the gas flows, electrically controlled needle valves adjust and maintain the inlet pressure with the assistance of a pressure gauge (capacitance gauges) mounted on the room temperature part of each line. The inner dimensions of the main body of the buffer-gas cell, which is mounted to the bottom plate of the cryostat, are $2.5 \times 2.4 \times (l_{cell}=1)\,$cm$^3$. To optimize performance the cell length $l_{cell}$ can be varied. The buffer-gas cell can be shifted back and forth to allow changes of $l_{cell}$ without altering the guide segment directly after the buffer-gas cell. To extend the cell length, additional $1\,$cm long extensions can be inserted. The exit aperture is formed by a $1\,$mm hole in a copper foil, which is glued on a Kapton foil to thermally disconnect it from the cell (see Section 5.3). This Kapton foil is clamped on the exit plate of the cell, which has a $10\,$mm opening.

\subsection{Electric Guide}
\label{electrodes}
\begin{figure}[t]
\begin{center}
\includegraphics[scale=0.9]{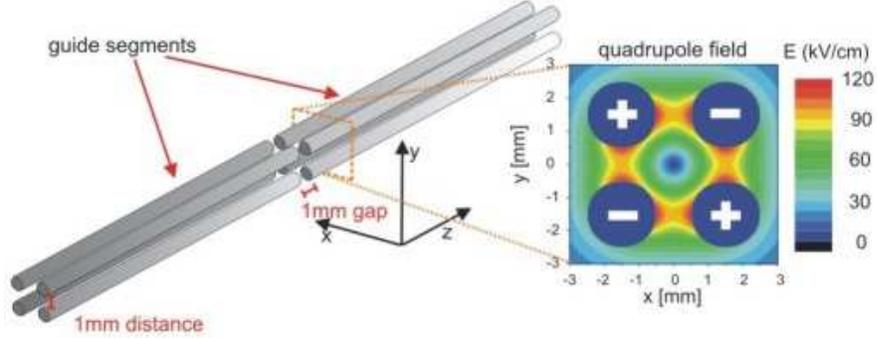}
\caption{Electrode arrangement. A guide segment consists of four stainless steel rods, each with a diameter of $2\,$mm, located at a distance of $1\,$mm to a neighboring electrode in a quadrupole configuration. The $1\,$mm gap between the guide segments keeps the guided molecules largely unaffected. The electric field distribution for $\pm 5\,$kV on the electrodes is shown for a cross sectional area perpendicular to the propagation direction of the molecules.}
\label{fig4}
\end{center}
\end{figure}

To extract the molecules from the buffer-gas environment and obtain a pure sample of cold molecules, a bent electrostatic guide segment is placed in front of the exit aperture of the buffer-gas cell as shown in Fig.~\ref{fig3}. 
The distance between the guide and the buffer-gas cell is around $1\,$mm to prevent discharges. A guide segment is made of four stainless steel rods, each with a diameter of $2\,$mm \cite{Junglen2004a}. These electrodes are arranged in a way that a two-dimensional quadrupole field configuration is formed when positive and negative voltages are applied such that neighboring electrodes, separated by $1\,$mm, have opposite polarity (see Fig.~\ref{fig4}). The electrodes are held by metallic holders which are mounted on an insulating ceramic material (Macor). High voltage is applied via Kapton insulated wires, connected to the holders. A bend of $90^{\circ}$ with a $2.5\,$cm radius of curvature is made in the first quadrupole guide segment after less than $2\,$cm from the exit of the buffer-gas cell. The total guide consists of four quadrupole segments, separated by $1\,$mm gaps. Two segments are located in the vacuum vessel with a length of $12\,$cm for the first and $31.4\,$cm for the second segment. Two other segments with a length of $36.6\,$cm and $26.8\,$cm are in the high-vacuum region, placed above the vacuum vessel of the cryogenic source. These segments are used to deliver the guided beam to the detector. The last straight segment allows for depletion spectroscopy by overlapping the central axis of the guide with a laser beam \cite{vanBuuren2008}.

The high-vacuum region can be separated from the vessel by a valve. The valve consists of a stainless steel mechanical shutter which can be pushed through the $1\,$mm gap between the second and third guide segment. From below it is sealed by a Viton O-ring. The valve is used to separate the high-vacuum region from the vessel during warm up of the cryogenic part of the set-up. In this way a vacuum better than $\sim 10^{-8}\,$mbar can be maintained, even when an atmospheric pressure is applied inside the vessel. This enables us to adjust the source without polluting the high-vacuum region. To preserve a low background pressure ($< 10^{-9}\,$mbar) in the detection chamber during measurements, when large amounts of helium gas are flown into the buffer-gas cell, we use two differential pumping stages. The first stage is placed above the mechanical valve and the other between the two cubes inside the high-vacuum region (see Fig.~\ref{fig2}). The differential pumping sections consist of two Macor brackets which are clamped directly to the guide rods via metallic holders. It allows molecules to travel between the different vacuum regions only through the small opening in the middle of the quadrupole guide. To maintain a low pressure in the high-vacuum region, we use three turbo molecular pumps (two with pumping speeds of $\sim 60\,$l/s and one with a pumping speed of $\sim 210\,$l/s for the detection chamber) backed by another turbo pump and membrane pump. Pressures of $10^{-9}\,$mbar in the region behind the valve and $10^{-10}\,$mbar in the detection chamber can be obtained. Ion pressure gauges are installed in the vessel and near the detection unit.

For the detection of guided molecules we use a quadrupole mass spectrometer (QMS) with a crossbeam ion source, located in the detection chamber (PFEIFFER QMA $410$). This analyzer provides an ion counter, which is used to determine the
flux and density of the guided gas. The mass spectrometer is mounted on a translation stage to change the position of the ionization volume relative to the guide exit.

\section{Data Acquisition}
\label{data}
To count the guided molecules with the QMS and to distinguish their signal from the background, we switch the high voltage on the guide on and off repeatedly. The difference in count rate reflects the contribution of guided molecules.
Velocity distributions can be obtained from the time-of-flight (TOF) signal after switching on the high-voltage, as described in \cite{Junglen2004a}. The first and second guide segments are connected to high-voltage transistor push-pull switches (BEHLKE HTS $151$-$03$-GSM), each housed in a copper box to prevent the generated radio frequency from being radiated. TTL pulses from a pulse generator are applied to the switches to set the on ($T_{ON}=110\,$ms) and off ($T_{OFF}=100\,$ms) configurations. This timing scheme allows detection of molecular velocities down to about $10\,$m/s, well within the regime where noise and systematics become dominant. This is comparable to the effects found in our previous guiding experiments \cite{Junglen2004a}. The third and fourth segment are permanently on high voltage to avoid pick-up currents on the QMS electrodes. 

The QMS ion counter signal is sent to a TTL pulse shaper that generates $\sim 100\,$ns TTL pulses if the amplitude of the incoming count signal exceeds a fixed threshold voltage. The threshold is used to eliminate noise. The TTL pulses are recorded by a MCS (Multi Channel Scalar) card, which creates histograms of the counts as a function of time. The trigger for the scalar card is generated by the pulse generator and starts $10\,$ms before the high voltage is switched on. The data from the temperature and pressure sensors are recorded with a LabVIEW interface.

\section{Flux Calibration}
\label{calibration}
To determine the flux of guided molecules, the sensitivity of the QMS for the guided species has to be determined. To calibrate the QMS, we apply a constant flow of molecular gas (of the same species as the molecules that have been guided) to the detection chamber through a needle valve, mounted on a side flange. The density inside the chamber is determined from the pressure measured by an ion gauge (VARIAN Bayard/Alpert gauge). When the pressure has stabilized, we take a mass spectrum ($1-100\,$amu) to determine the masses that contributed to the increase in pressure as well as the distribution of ionization fragments. The next step is to measure the count rate at the particular mass of the guided molecules. This is done for various densities in the detection chamber. For the mass spectrometer as well as the ion gauges the ionization probability of the investigated molecules has to be taken into account. Since the mass spectrometer is operated in a regime far from saturation for our guided molecules, the QMS signal is proportional to the density of the molecules in its ionization volume. We checked by variation of the ionization current that the ionization is not saturated, not even for slow molecules.

\begin{figure}[t]
\begin{center}
\includegraphics[scale=0.9]{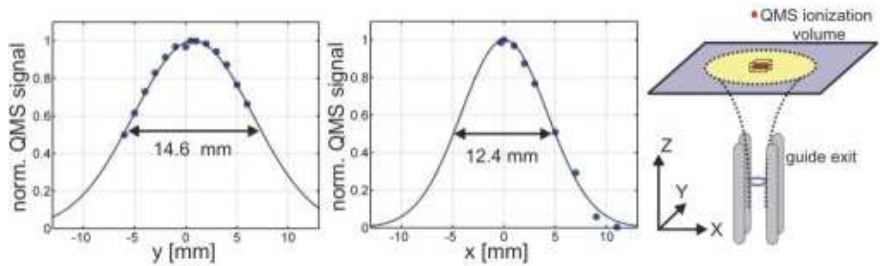}
\caption{Measurement of the beam shape in (y) and (x) direction fitted by a Gaussian distribution. The distance between the guide exit and the ionization volume of the mass spectrometer is $\sim 25\,$mm along the z-direction. The profile in the y-direction is widened compared to the x-direction due to the centrifugal distortion of the trap potential in the bent path of the guide. For the beam in the guide a Gaussian profile is assumed with a standard deviation of $\sigma = 400\,\mu$m, based on simulations presented in \cite{Junglenthesis}.}
\label{fig5}
\end{center}
\end{figure}

Since the mass spectrometer detects only molecules that enter the ionization volume, we have to determine the fraction of the guided beam reaching the QMS ionization volume. By varying the position of the QMS ionization volume in the plane perpendicular to the guide exit, the beam spread is obtained (see Fig.~\ref{fig5}). Since the width of the ionization volume is much smaller than the expanded beam, the beam shape can be resolved. The beam profile in the plane is well described by a Gaussian distribution. With this distribution the ratio of undetected to detected molecules is determined by a simple integration of the normalized Gaussian over its peak, with the integration boundaries given by the ionization region. Using this factor, the average velocity of the guided molecules ($65\,$m/s)\footnote[2]{The average velocity is obtained from a velocity distribution measurement of the guided molecules.} and the measured density, total fluxes of guided molecules of $\sim 10^{10}\,$s$^{-1}-10^{11}\,$s$^{-1}$ are obtained. Assuming a Gaussian profile of the beam in the guide, as supported by numerical simulations \cite{Junglenthesis}, we obtain a peak density in the guide of $\sim 10^{9}\,$cm$^{-3}$ \cite{vanBuuren2008}.

\section{Optimization of the source}
\label{optim}
To maximize the flux of slow molecules and to improve the cooling process, several parameters of the source are optimized. Different cell lengths, gas densities and exit apertures are tested in our set-up. In the end a compromise had to be found since optimal parameters for the flux, internal cooling and runtime of the source do not coincide perfectly with each other.

\subsection{Flux Optimization}
\label{fluxmax}
\begin{figure}[t]
\begin{center}
\includegraphics[scale=0.8]{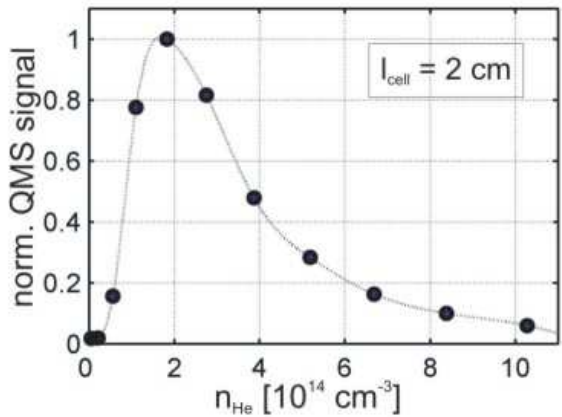}
\caption{Normalized QMS signal as a function of the helium density in a $2\,$cm cell with a fixed ND$_{3}$ density of $\sim 1.4 \times 10^{16}\,$cm$^{-3}$ at the input of the cell. The buffer-gas density is varied over about one order of magnitude. The rising edge indicates the region of insufficient cooling. The falling edge displays the effect of boosting as discussed in the text. At the peak position of $n_{\rm He}$ a flux of order of $10^{10}\,$s$^{-1}-10^{11}\,$s$^{-1}$ is obtained \cite{vanBuuren2008}. The curve connecting the data points is a guide to the eye.}
\label{fig6}
\end{center}
\end{figure}

In this section, the optimization of the flux with respect to the buffer-gas and molecular-gas densities is presented. Most of the discussed measurements are performed with a $2\,$cm long cell, although we also investigated other cell lengths. In Fig.~\ref{fig6} the normalized QMS signal is shown for different buffer-gas densities. The characteristics of this curve can be explained in the following way. For low helium densities most of the molecules are still too fast and can escape from the guiding field of the bent quadrupole. With higher buffer-gas densities the cooling increases, and the signal rises to the maximum. If the density is raised to higher values the signal begins to drop. This is explained by an effect that is called "boosting" \cite{Maxwell2005} and which is described in the following. Since the density of helium near the cell exit is higher than the molecular gas density at that position, molecules are most likely to collide with helium atoms. In this case, the probability for collisions between helium atoms and molecules, represented by the mean free path $\lambda_{mean} = 1/(n_{\rm He}\times \sigma)$, predominantly depends on the density of the helium atoms $n_{\rm He}$, where $\sigma$ is the helium-molecular (elastic) cross section. If the exit aperture dimension ($1\,$mm diameter) is larger than the mean free path, there is a high probability for molecules leaving the cell to get accelerated (boosted) in the forward direction. This happens due to collisions with helium atoms approaching the molecules from behind because of the pressure gradient and the large average velocity of helium ($\langle v_{He} \rangle \sim 230\,$m/s at $5\,$K), much larger than the velocity of the molecules.

\begin{figure}[t]
\begin{center}
\includegraphics[scale=0.9]{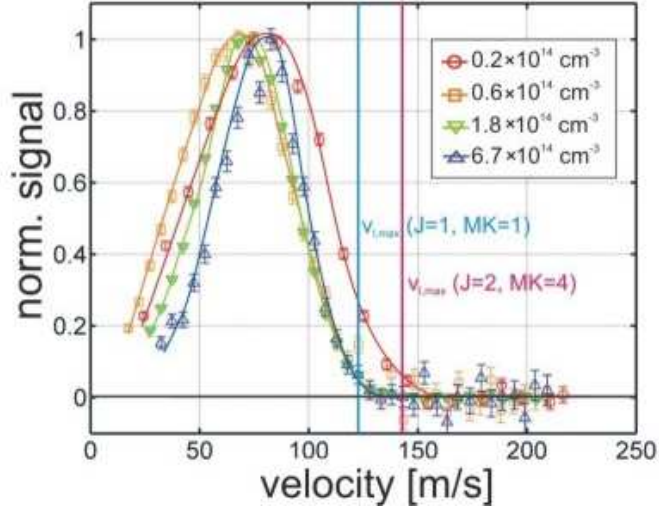}
\caption{Normalized velocity distribution for a $2\,$cm cell configuration with different helium densities. The (ND$_3$) molecular density at the cell entrance is fixed to $\sim 1.4 \times 10^{16}\,$cm$^{-3}$. The cut-off velocities for the lowest-lying low-field-seeking rotation state $(J=1,MK=1)$ and for the first higher-lying low-field-seeking state $(J=2,MK=4)$ are marked by the straight lines. The data taken at the lowest density ($0.2\times 10^{14}\,$cm$^{-3}$) show contributions of velocities higher than the cut-off velocity for the $(J=1, MK=1)$ state. This indicates the presence of higher-lying rotational states and, therefore, insufficient rotational cooling. Compared to this, the curve at $0.6\times 10^{14}\,$cm$^{-3}$ is shifted to lower velocities, indicating better cooling. The identical high-velocity sides of the highest two densities indicate that the internal state distribution has stabilized. However, the high helium densities above $0.6\times 10^{14}\,$cm$^{-3}$ do lead to 'boosting', the shift of the rising edge to higher velocities. The velocity distributions are obtained from TOF signals which are cut off if $99\,$\% of the steady-state value is reached, before noise and systematics become too dominant. The curves connecting the data points are guides to the eye.}
\label{fig7}
\end{center}
\end{figure}

\begin{figure}[t]
\begin{center}
\includegraphics[scale=0.8]{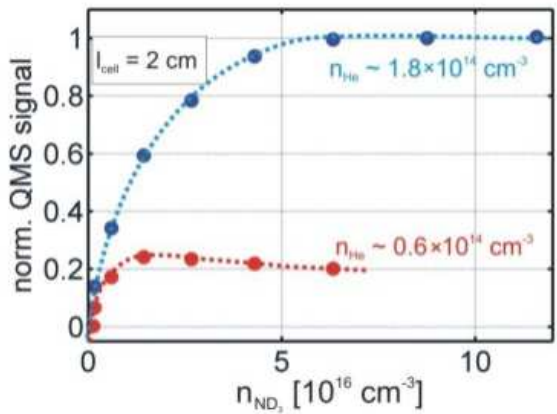}
\caption{Optimization of the ammonia density in the buffer-gas cell. The values for the ammonia density are given for the inlet tube exit. Since most of the ammonia is pumped away by the cell walls, the density close to the exit of the buffer-gas cell will be much less than its original value at the inlet. The helium density is fixed to $0.6\times 10^{14}\,$cm$^{-3}$ and $1.8\times 10^{14}\,$cm$^{-3}$, the latter being the optimum value derived from the buffer-gas scan. In both cases the QMS signal reaches an optimum. For the low buffer-gas density case, we see that the signal starts to drop for higher ammonia densities, which might be explained by insufficient cooling power of the helium to cool down the molecules. For the higher buffer-gas density the signal saturates, whereby a signal drop is not visible even for much higher ND$_{3}$ densities. For all other measurements the ammonia density was set to a value of $\sim 1.4 \times 10^{16}\,$cm$^{-3}$. At this setting a lower flux is obtained, but it allows for longer measuring times due to less ice formation. The dashed curves are guides to the eye.}
\label{fig8}
\end{center}
\end{figure}

The effect of boosting is visible in velocity distributions obtained with high buffer-gas density. For very low buffer-gas densities ($0.2\times 10^{14}\,$cm$^{-3}$) the cooling is not optimal and can be improved (see Fig.~\ref{fig7}). This can be seen from the peak position and the cut-off velocity, defined as the velocity where the measured velocity distribution is compatible with zero. Both shift to lower velocities when the helium density is increased to $0.6\times 10^{14}\,$cm$^{-3}$. Information about the cooling for the internal degrees of freedom is obtained from the falling edge of the distribution at high velocities. This will be described in more detail in the upcoming Subsection \ref{internal}. For helium densities higher than $0.6\times 10^{14}\,$cm$^{-3}$, where collisions start to alter the shape of the velocity distribution, the rising edge is shifted to higher velocities, thereby supporting the argument of boosting.
From the ammonia density scans at fixed helium densities, in which various amounts of ND$_{3}$ are injected into the cell (Fig.~\ref{fig8}), we can see that the molecular flux saturates. We believe that the saturation comes from the limited heat capacity of the helium gas. If more ammonia is injected, each molecule is simply cooled less. This explanation is supported by the fact that a larger helium density can support a larger ND$_{3}$ flux at a larger saturation density.

Similar examinations performed with a $1\,$cm and $3\,$cm long cell length indicate that the maximal flux is obtained in a $2\,$cm long cell. This can be explained in the following way. In a shorter cell less collisions take place to cool down the molecules for equal buffer-gas densities. Therefore, higher buffer-gas densities have to be applied to a $1\,$cm long cell to obtain a maximal flux, in comparison to a $2\,$cm long cell. However, with a higher helium density the boosting is increased, which reduces the flux. For a $1\,$cm cell length the flux is by a factor of two lower than in the $2\,$cm configuration. If the cell is extended to $3\,$cm, we get a reduction of the flux due to the smaller ratio of exit aperture to inner surface of the cell. This means that less molecules find their way to the exit and a large fraction of them get stuck on the walls of the cell at cryogenic temperatures.
We come to the conclusion that the highest flux is obtained for a buffer-gas density of $n_{\rm He} \sim 2 \times 10^{14}\,$cm$^{-3}$ and an ammonia density of $n_{\rm ND_{3}} \sim 5\times 10^{16}\,$cm$^{-3}$ in the inlet of a $2\,$cm long cell. These parameter settings are throughout the rest of this paper called the optimal flux settings, which can be different for other molecules. For most of the measurements we used $n_{\rm ND_{3}} \sim 1.4 \times 10^{16}\,$cm$^{-3}$ to have longer stable running periods as discussed in Subsection \ref{runtime}.

\subsection{Optimization of Internal Cooling}
\label{internal}
From the velocity distributions for different buffer-gas densities we can extract information about the internal cooling of the molecules. This is possible because the cut-off velocities in the longitudinal direction (highest velocity that is populated in the velocity distribution) depends mainly on the radius of curvature of the bend in the guide ($R = 2.5\,$cm) and the state-dependent Stark shift of the molecules. The Stark shift of ammonia is given by the approximation
\begin{equation}\label{ND3Stark}
\Delta W^{s} = \pm \sqrt{(W_{inv}/2)^{2} + (\mu|\vec{E}|\frac{MK}{J(J+1)})^2},
\end{equation}
which is valid for electric fields $|\vec{E}|$ in the range of $0\,$kV/cm$\,-\,100\,$kV/cm \cite{Schawlow}. The dipole moment of ammonia $\mu = 1.5\,$D, $J$ represents the main rotation quantum number with its projections $K$ on the molecular symmetry axis and $M$ on the axis of the external electric field, and $W_{inv} = 0.053\,$cm$^{-1}$ is the inversion splitting of ND$_{3}$. We can determine the cut-off velocities
\begin{equation}
v_{l,max} = \sqrt{\Delta W^{s}(E_{max})\frac{R}{r\cdot m}}
\label{cutoff}
\end{equation}

\cite{Junglen2004a}, with the inner guide radius $r = 1.12\,$mm specifying the location of the electric field maximum $E_{max}\sim 90\,$kV/cm for $\pm 5$ kV electrode voltage (see Fig.~\ref{fig4}) and the mass of the molecule $m = 20\,$amu. The cut-off velocity is a marker for the state with the highest Stark shift of the guided molecules. To obtain information on the cooling process for the internal degrees of freedom, we compare the calculated cut-off velocities (Table~\ref{tab1}) with measured velocity distributions (see Fig.~\ref{fig7}). For low temperatures, large populations in low-lying rotational states are expected. As can be seen from Fig.~\ref{fig7}, the fraction of the normalized distributions above the cut-off velocity of the lowest-lying guidable rotational state ($J=1$, $M K =1$) is reduced if the helium density is increased above $0.2\times 10^{14}\,$cm$^{-3}$. If the helium density is increased above $0.6\times 10^{14}\,$cm$^{-3}$ the cut-off velocity does not seem to change anymore. Therefore we come to the conclusion that the population of higher rotational states and especially of those with a higher Stark shift than $(J=1 ,M K =1)$ is negligibly small in these cases.
\begin{table}
\begin{center}
\caption{
Lowest lying low-field-seeking rotational states of ND$_{3}$ molecules up to $J=2$. The Stark shifts of the rotation states are calculated with equation (1) and (2) for an electric field strength of $\sim 90\,$kV/cm, which corresponds to $\pm 5\,$kV on the electrodes. The maximal longitudinal velocities are evaluated for the same settings. The rotational constants of ammonia, which are needed to calculate the rotational energies, are taken from \cite{Herzberg}. }

\begin{tabular}{lll|llll}
\hline
$J$ &$|K|$ &$|M|$ &E$_{rot}$ (cm$^{-1}$)  &Stark energy  (cm$^{-1}$)  &v$_{l,max}$
 (m/s)\\
\hline
1   &1  &1   &8.29    &1.13   &123\\
2   &1  &1   &28.85   &0.38   &71\\
2   &1  &2   &28.85   &0.76   &100\\
2   &2  &1   &22.88   &0.76   &100\\
2   &2  &2   &22.88   &1.51   &142\\
\hline
\end{tabular}
\end{center}
\label{tab1}
\end{table}

In summary, the velocity distributions presented in Fig.~\ref{fig7} indicate that the molecules are rotationally cooled in the buffer-gas environment of a $2\,$cm set-up. Since this is just a qualitative measure of internal cooling, we established a depletion spectroscopy set-up, as described in \cite{Motsch2007}, for the cryogenic source. In this experiment, we obtained quantitative results for the population of internal states of formaldehyde. For this molecular species a $\sim 80\,$\% pure beam was obtained at optimal flux setting of $n_{\rm He}$ \cite{vanBuuren2008}.
The disadvantage of the depletion method is, however, that different laser frequencies and possibly even different laser systems are required for different molecular species. The measurement of the cutoff velocity, in contrast, is simple and can readily be performed for any guidable molecule.

\subsection{Runtime Optimization}
\label{runtime}
\begin{figure}[t]
\begin{center}
\includegraphics[scale=0.9]{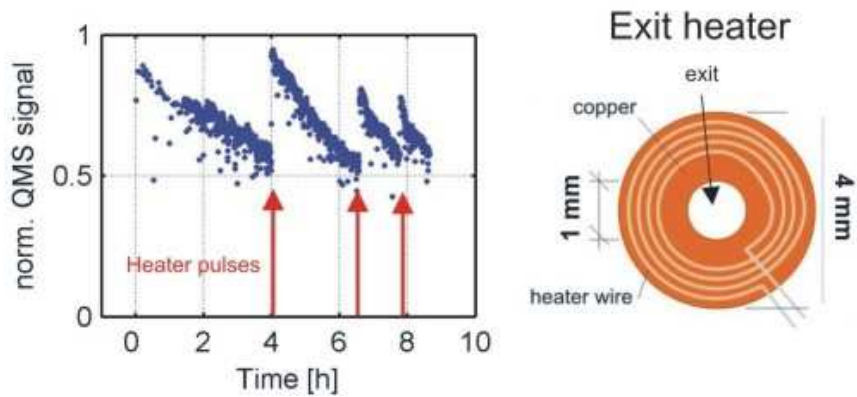}
\caption{Left: ND$_3$ guiding signal as a function of time showing the decrease in signal due to ice formation for a period of $\sim 9\,$hours and three de-icing heater pulses. Right: a sketch of the heater plate, which is placed on the exit aperture of the buffer-gas cell. The plate consists of a $0.06\,$mm thin copper foil with a $0.22\,$mm thick heater wire attached to it. The signal can be restored several times, when heater pulses with $\sim 0.3\,$W of $\sim 1\,$min duration are applied.}
\label{fig9}
\end{center}
\end{figure}

The measurement time in our set-up is limited by the sticking of molecular gas to the walls of the buffer-gas cell, which happens long before the saturation of the charcoal by helium gas. The charcoal can pump helium gas for several days in the density regimes we use in our system. However, we observe a steady decrease of the flux of guided molecules with a half time of $\sim 5\,$hours, which we attribute to frozen molecules clogging the entrance and the exit aperture of the cell. The ice formation has been observed directly after the installation of viewing ports in the vacuum vessel and radiation shields. Especially in the exit aperture large blocks of molecular ice appear after several hours of operation; for a smaller opening of $\sim 0.5\,$mm instead of $1\,$mm this is barely $30\,$min. The ice reduces the diameter of the exit aperture and with it the amount of molecules that can enter the guide. Since the warm up of the complete set-up takes $\sim 8\,$hours, a different solution had to be found.

We use a Nichrome heater wire ($10\,\Omega$/m), attached with cyanoacrylate based fast-acting glue to a $4\,$mm diameter copper foil that has a $1\,$mm opening in the middle. The heater plate is glued to a Kapton foil with the wires pointing into the cell. The coil is additionally covered with Stycast $2850$ epoxy for better adherence and thermal connection. A Pt$100$ temperature sensor is glued to the copper plate as well. It is needed to carefully monitor the temperature. Since the heater is glued to the Kapton foil and has a very low thermal contact with the copper frame we can easily alter the temperature of the exit aperture.
From Fig.~\ref{fig9} we see that the signal indeed recovers each time a heater pulse is applied to the exit aperture. However, we see that the measurement periods after recovery get smaller. We attribute this to the growing ice layer within the cell. After $9\,$hours of measurement with optimal flux setting of the helium density and an ammonia density of $\sim 1.4 \times 10^{16}\,$cm$^{-3}$ at the inlet of the cell, we approach the time limit for continuous operation. At this ammonia density we then have $\sim 50\,$\% of the maximum flux (see Fig.~\ref{fig8}). Of course we could run at higher ammonia densities, but at the cost of shorter running times due to faster ice formation. After warming up the whole cell to release the frozen molecular gas, the initial flux can be fully recovered and the system is ready again for operation.

\section{Cold Beams of Different Molecular Species}
\label{species}
Most of our measurements are performed with deuterated ammonia (ND$_{3}$) and formaldehyde (H$_{2}$CO)
because these molecules have been used in our previous room-temperature experiments \cite{Junglen2004a,Motsch2007} and because ND$_3$ has also been employed in buffer-gas cooling experiments \cite{Willey2004}. Besides, as already mentioned in section \ref{internal}, H$_{2}$CO has transitions in the near ultraviolet, with which we have quantified the purity of the produced beam using depletion spectroscopy \cite{vanBuuren2008}. These data gave direct evidence of internal cooling of the molecules by collisions with the buffer gas. To demonstrate that our cryogenic source is a versatile tool for generating cold molecules, we also investigated the production of cold guided beams consisting of other symmetric top molecules such as trifluoromethane (CF$_{3}$H) and fluoromethane (CH$_{3}$F). The masses, dipole moments and cut-off velocities of the lowest guidable state ($J=1$, $M K=1$), calculated with Eq. \ref{cutoff}, for the molecules employed in this work are listed in Table~\ref{tab2}.

\begin{figure}[t]
\begin{center}
\includegraphics[scale=0.9]{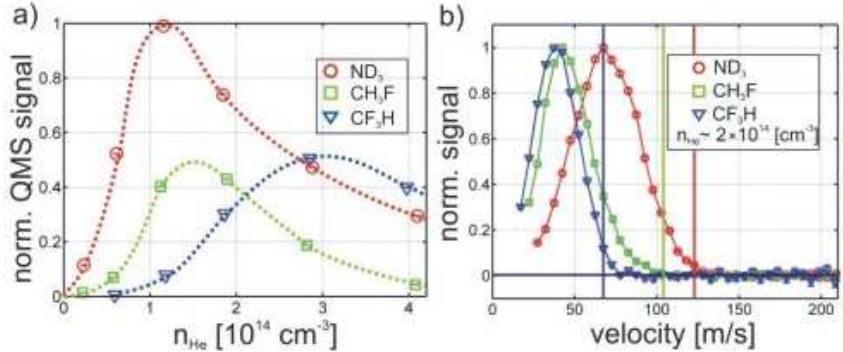}
\caption{(a) Buffer-gas scans for ND$_{3}$, CH$_{3}$F and CF$_{3}$H, normalized to the peak value of the ND$_{3}$ curve. For higher masses the optimal value of the buffer-gas scan is shifted to higher densities since more collisions are needed to slow down the heavier molecules. The dashed curves in (a) are guides to the eye. In (b) the velocity distributions for the three different species at a similar buffer-gas density ($n_{\rm He} \sim 2\times 10^{14}\,$cm$^{-3}$) are compared. The cut-off velocities of the distributions are shifted to lower velocities for higher masses. The vertical lines indicate the calculated positions of the cut-off velocities for the $|J=1,MK=1\rangle$ low-field seeking state.}
\label{fig10}
\end{center}
\end{figure}

The measurements with CF$_{3}$H and CH$_{3}$F are performed without any change to the source except for the gas handling system. Due to the higher vapor pressure of these molecules, the molecular gas line was only heated to $\textrm{T}\sim 140\,$K instead of $\textrm{T}\geq 180\,$K for ND$_3$. In Fig.~\ref{fig10}a buffer-gas scans are presented for the different species. To cool down heavy molecules, more collisions with helium are needed than for light molecules. Therefore, the position of the peak of the buffer-gas scan for the heavier CF$_{3}$H is shifted to higher $n_{\rm He}$ as compared to the other lighter molecules. The QMS signals obtained for CF$_{3}$H and CH$_{3}$F are only a factor of two lower than the ND$_3$ signal. Since we calibrated the QMS only for ND$_3$, we did not determine absolute numbers for the guided fluxes of the other species. However, we can conclude from these high signals that fluxes and densities comparable to those for ND$_3$ are feasible with this set-up, showing the universality of our source.
In these measurements, we did not compensate for the change in the conductance of the molecular gas line due to the different molecular masses and different gas line temperatures. This change causes the molecular densities in the inlet to vary for the different data sets, but we expect no significant effects on the results. Note the shift in the peak position of the buffer-gas scan of ND$_{3}$ in Fig.~\ref{fig10}a in comparison to Fig.~\ref{fig6}. This probably results from a rebuild of the buffer-gas cell between these two measurements. As observed from several rebuilds, the peak position can shift by $\sim (\pm 1)\times 10^{14}\,$cm$^{-3}$ after such an intervention. Therefore, buffer-gas scans are performed after each rebuild of the system to relocate the optimal setting for $n_{\rm He}$.

The velocity distributions of the three different species are shown in Fig.~\ref{fig10}b for a fixed helium density of $n_{\rm He}$ = $2 \times 10^{14}\,$cm$^{-3}$. The vertical lines indicate the cut-off velocities of the lowest guidable state of each molecule. The velocity distributions of both ND$_3$ and CH$_{3}$F are compatible with zero at the cut-off velocities of the lowest guidable state. This is not the case for CF$_{3}$H as discussed later. Since $\Delta W^s$ for states with $J^2=M  K$ increases with $J$ for symmetric top molecules in first order according to \cite{Schawlow},
\begin{equation}
\Delta W^s=\mu E\frac{MK}{J(J+1)},
\label{Ws}
\end{equation}

\begin{figure}[t]
\begin{center}
\includegraphics[scale=0.9]{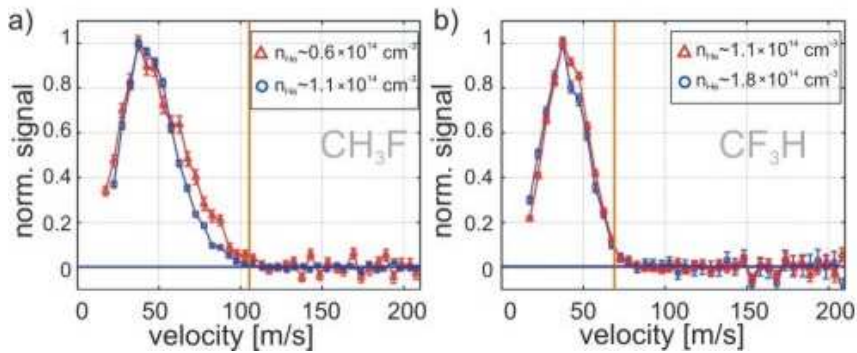}
\caption{Velocity distributions of CH$_{3}$F (a) and CF$_{3}$H (b), normalized to the maximum value. Helium densities resulting in maximum flux and below the optimal value are used to illustrate the effect of cooling. In case of CH$_{3}$F the data taken at low $n_{\rm He} \sim 0.6 \times 10^{14}\,$cm$^{-3}$ contain a marginally smaller boosting than the data taken at the optimal flux settings of $n_{\rm He} \sim 1.1 \times 10^{14}\,$cm$^{-3}$. Even with this smaller boosting, the low $n_{\rm He}$ data show higher contributions at high velocities. This indicates population in higher rotational states for which the cut-off velocities are higher than for the $J=1, K=1, M=1$ state. The vertical lines indicate the calculated positions of the cut-off velocities for the $(J=1,M K=1)$ low-field seeking state. A similar behavior is shown for ND$_3$ in Fig.~\ref{fig7}. For CF$_{3}$H we do not observe such differences between velocity distributions taken at different $n_{\rm He}$, except for a very small difference in boosting. This can be explained by the small rotational constants of CF$_{3}$H, causing populations in higher states even at temperatures around 5 K.
}
\label{fig11}
\end{center}
\end{figure}
the CH$_{3}$F and ND$_{3}$ data (for which Eq. \ref{ND3Stark} is valid) 
indicate that the beams consist mainly of states with $J=1$. Large populations in higher $J$ states would show higher cut-off velocities due to contributions of states with $J^2=M  K > 1$. In other words, Fig. \ref{fig10}b shows that the ND$_3$ and CH$_3$F molecules are rotationally cooled by collisions with the buffer gas. For ND$_3$, the onset of rotational cooling is demonstrated in Fig.~\ref{fig7}b, by comparing the shapes and cut-off velocities of velocity distributions taken at different buffer-gas densities.

We also performed measurements at different buffer-gas densities for CF$_{3}$H and CH$_{3}$F. Fig. \ref{fig11} shows two velocity distributions for both CH$_{3}$F and CF$_{3}$H. One distribution is taken at the optimal flux setting of $n_{\rm He}$ and the other for a lower value of $n_{\rm He}$. The data taken at low $n_{\rm He}$ display a marginally smaller boosting than the data taken at the optimal flux settings for both gases. Although the low $n_{\rm He}$ data of CH$_{3}$F exhibit smaller boosting, the data show larger contributions at high velocities. This is explained by population in higher rotational states for which the cut-off velocities are larger than for the $(J=1, MK=1)$ state. We have shown in \cite{vanBuuren2008} that the internal cooling for H$_2$CO is already near its maximal value at the optimal flux setting of $n_{\rm He}$ and also the velocity distributions of ND$_3$ and CH$_3$F indicate qualitatively that most of the molecules are in the lowest guidable state ($J=1$, $MK=1$) at this setting. Therefore, we could expect that the CF$_3$H molecules are rotationally cold at their optimal flux setting of $n_{\rm He}$ as well. However, comparing data taken at different buffer-gas densities does not reveal any effect of internal cooling as shown in Fig. \ref{fig11}b. This can be explained by the small rotational constants of CF$_{3}$H ($A_0 = 0.19\,$cm$^{-1}$ and $B_0=0.35\,$cm$^{-1}$ \cite{Wensink1979}), causing population in higher states even at temperatures around 5 K. Since $\Delta W^s$ converges to $\mu E$ for high $J$ states with $J^2=MK$ (see Eq. \ref{Ws}), a distinct shift towards smaller cut-off velocities is only visible between molecules populating mainly $J=1$ states and molecules populating states with higher $J$. Therefore, we conclude that although the guided beam of CF$_{3}$H is not pure, it is likely to consist of a few states only with low rotational quantum numbers and low rotational energy. To increase the purity even more, one could deplete higher rotational states if light for the correct transitions is available. This would, however, cause an overall reduction of the flux. A better option would be to reduce the temperature of the buffer gas even further. For CF$_3$H temperatures around 1 K would suffice to increase the purity strongly. The cooling of CaH and CaF molecules in buffer gases with temperatures below $1\,$K has already been demonstrated \cite{Weinstein1998, Maussang2005}.
\begin{table}
\begin{center}
\caption{
The masses, dipole moments and maximal longitudinal velocities of deuterated ammonia ND$_{3}$, fluoromethane CH$_{3}$F and trifluoromethane CF$_{3}$H.}
\begin{tabular}{l|lll}
\hline
\\
  & mass (amu) &dipole moment (D) &v$^{(J=1, |MK|=1)}_{l,max}$
 (m/s) \\
\hline
ND$_{3}$    &20  &1.5   &123 \\
CH$_{3}$F   &34  &1.86  &105 \\
CF$_{3}$H   &70  &1.65  &69 \\
\hline
\end{tabular}
\end{center}
\label{tab2}
\end{table}

In summary, we were able to produce cold guided beams of several different molecular species. Our data suggest that these beams consist of molecules in low rotational state(s), where the number of populated states depends on the density of rotational states at low temperature.

\section{Conclusion and Outlook}
\label{summary}
With our cryogenic source we produce continuous guided beams of slow and internally cold polar molecules. The characteristics of the source are examined with respect to variations in the molecular density, helium buffer gas density and cell length. From these studies, optimized settings for the system are found to produce maximal fluxes. Besides, signatures of internal cooling are obtained from the velocity distributions taken at different buffer-gas densities. The latter has been confirmed by depletion spectroscopy measurements on the guided beam emitted from the cryogenic source \cite{vanBuuren2008}. By comparing the cooling and guiding characteristics for four different molecular gases, we showed that our source is applicable to a wide variety of molecular species.

With the current set-up further experiments can be carried out. For example, molecules in the ro-vibrational ground state could be guided by applying alternating electric fields to the guide \cite{Junglen2004}. In this experiment one could use depletion spectroscopy for detection. The laser depletion technique itself could be used as a tool to further purify the guided beam. To increase the guided flux in an electrostatic quadrupole guide, Raman pulses or microwave fields could pump molecules from the high-field-seeking, and therefore not guidable ground state, into the first excited low-field-seeking state. This could be done in the small gap between the exit of the cell and the first guide segment.
With an extension to a large-volume electric trap \cite{Rieger2005}, collision studies could be realized. Slow molecules from the same or different species could be brought into collision under controlled conditions \cite{Willitsch2008}. With the ability to choose the rotational state of the molecules from our source, it would be possible to measure state-selective cross sections.
Our cryogenic source could also be used as a new starting point for further cooling schemes.

\appendix

\section{Gas Line Calibration}
\label{lines}
{In this appendix, we present the calibration of the helium and molecular gas lines. The calibrations are used to convert values from the room-temperature pressure gauges into densities in the buffer-gas cell. The measurement} is done at room temperature with nitrogen (instead of helium, ammonia or other molecular gases) since the pumping efficiency of the turbo pump (on the bottom of the vessel) is specified by the manufacturer for nitrogen gas (S$_{pump}=550\,$l/s for N$_{2}$). We have removed the radiation shields and the buffer-gas cell in this experiment so that the conductance is mainly determined by the gas line itself.
The nitrogen gas is fed through one of the gas lines into the vacuum vessel and is pumped by the turbo molecular pump. A pressure gauge in the gas handling system, another one in the vessel and a mass spectrometer mounted on the side of the vessel are used for detection. The mass spectrometer is needed to determine which masses contribute to the rise in pressure. With the pressure detected in the vessel we can determine the flux into the chamber via $$\Phi = \frac{p_{vessel}S_{pump}}{k_{B}T_{room}},$$ where $p_{vessel}$ is the pressure in the vessel, $T_{room}=293\,$K and $k_{B}$ the Boltzmann constant. From this the conductance of the line can be obtained by taking into account the pressure in the gas handling system $p_{gas}$. This gives $C_{line} = \Phi k_{B}T_{room}/p_{gas}$. Next, correction factors for the different temperatures and masses must be applied. As a result the helium line conductance is $$C^{He}_{line} = C_{line} \sqrt{\frac{5 [\textrm{K}]}{295 [\textrm{K}]} \times \frac{28 [\textrm{amu}]}{4 [\textrm{amu}]}} .$$ Similar correction factors are required for the molecular gas line. With the conductance of the $1\,$mm aperture $C_{exit}$ estimated from \cite{Lafferty} $C_{exit}= A\sqrt{R_{0}T/(2\pi M_{m})}$ with $M_{m}$ the molar mass of molecular nitrogen used for calibration, the universal gas constant $R_{0}$ and the aperture area $A$ at $T = 5\,$K (valid for the molecular regime), we determine the values for the densities in the cell from $n_{cell}= \Phi/C_{exit}$.



\end{document}